\documentclass[prd,12pt]{revtex4}
\usepackage{slashed,graphicx,amsmath}

\begin{document}

    \title{Statistical Constraints on the Error of the Leptonic CP Violation of Neutrinos}
    \author{Kai Fu, Ying Zhang\footnote{E-mail:hepzhy@mail.xjtu.edu.cn. Corresponding author. }}
    \address{School of Science, Xi'an Jiaotong University, Xi'an, 710049, China}

\begin{abstract}
A constraint on the error of  leptonic CP violation, which require the phase $\delta_{CP}$ to be less than $\pi/4$  for it to be distinguishable on a $2\pi$ cycle, is presented. Under this constraint,
the effects of neutrino detector 's distance, beam energy, and energy resolution are discussed with reference to the present values of these parameters in experiments. Although an optimized detector performances can minimize the deviation to yield a larger distinguishable range of the leptonic CP phase on a $2\pi$ cycle, it is not possible to  determine an arbitrary leptonic CP phase in the range of $2\pi$ with the statistics from a single detector because of the existence of two singular points. An efficiency factor $\eta$ is defined to characterize  the distinguishable range of $\delta_{CP}$. To cover the entire possible $\delta_{CP}$ range, a combined efficiency factor $\eta^*$ corresponding to multiple sets of detection parameters with different neutrino beam energies and distances is proposed.  The combined efficiency factors $\eta^*$ of various major experiments  are also presented.

\vspace{0.5cm}
\end{abstract}
\vspace{1.5cm}

\maketitle

The neutrino CP phase, $\delta_{CP}$, is an important parameter in the  PMNS parameterization.
 It is related to neutrino CP violation effects and can be observed from Jarlskog invariant the $J_{CP}$ \cite{BrancoReview}.
 It also plays a  key role in astronomy and cosmology.
This  phase was predicted using several differently motivated models in \cite{NuCPVmodel1,NuCPVmodel2}.
Recently, a  $\delta_{CP}$ that is consistent with T2K data has been obtained from the Yukawaon model, which replaces the SM Yukawa couplings with the VEVs of Yukawaon scalars \cite{YukawaonModel}.
It has also been predicted using various newly proposed new mechanisms, such as the  $SU(3)$ gauge family symmetry proposed in \cite{SU3FamilySym} and the $\Delta(27)$ flavour symmetry proposed in \cite{Delta27Sym}.
Moreover, the leptonic CP phase has also been inferred through global fits to data from neutrino experiments  by various research groups \cite{GlobalFitOld1,GlobalFitOld2,GlobalFitOld3}. More recently updated optimized values can been found in Refs.\cite{GlobalFit20141,GlobalFit20142}.
However, to data, it has not been directly measured in experiments.
Future neutrino experiments including long-baseline facilities, superbeams and neutrino factories, are planned to measure the phase angle.
From the statistical point of view,  a central value of the phase alone is not sufficient to determine
its distribution on a cycle. Its error must also be considered.
In this paper, a constraint on the deviation of the leptonic CP violation angle is proposed based on statistics to ensure that it is determinate on a cycle.
Our motivation is to investigated the properties of this statistical constraint and its implications for neutrino experiments, especially reactor experiments.

The paper is organized as follows. Firstly, neutrino oscillation and decay probabilities, including the Earth matter effect, are reviews. The error on $\delta_{CP}$ is derived from the neutrino transformation probability. All contributions from  phenomenological parameters are investigated. Particular attention is paid to the  neutrino beam energy $E$ and detecting distance $L$ of the reactor from the reactor to achieve a suppression of the error on $\delta_{CP}$. The results require a set of $E$ and $L$ values that is  different from that required to achieve significant probability. A statistical constraint, i.e. the error on $\delta_{CP}$ must be less than $\pi/4$, is proposed as a requirement for the distinguishability of $\delta_{CP}$.  Then, we recommend a definition of efficiency factor $\eta$ to describe the range below $\pi/4$ on a $2\pi$ cycle of $\delta_{CP}$. The values of $\eta$ that correspond to current experiments are calculated, and the corresponding distinguishable  ranges of the leptonic CP phase are also shown. To obtain  full coverage of all possible $\delta_{CP}$ values, a combined $\eta$ based on multiple detectors is considered. Finally, a short summary is given.

 Let us review neutrino oscillations as a starting point. Neutrino oscillation experiments can be understood in terms of  rotations between the mass eigenstates $\nu_i$ ($i=1,2,3$) and the weak gauge eigenstates $\nu_\alpha$ ($\alpha=e,\mu,\tau$) as follows
    \begin{eqnarray*}
        \nu_\alpha=U_{\alpha i}\nu_i
    \end{eqnarray*}
Here, $U$ is the PMNS matrix and can be written in the form of $VK$. Generally, $V$ is parameterized as three 1-dimensional rotations
    \begin{eqnarray*}
        V=\left(\begin{array}{ccc}1&0&0\\
            0&c_{23}&s_{23}\\
            0&-s_{23}&c_{23}
            \end{array}\right)
            \left(\begin{array}{ccc}c_{13}&0&s_{13}e^{-i\delta_{CP}}\\
            0&1&0\\
            -s_{13}e^{i\delta_{CP}}&0&c_{13}
            \end{array}\right)
            \left(\begin{array}{ccc}c_{12}&s_{12}&0\\
            -s_{12}&c_{12}&0\\
            0&0&1
            \end{array}\right)
    \end{eqnarray*}
with the Majorana phase $K=\text{diag}(1,e^{i\delta_1},e^{i\delta_2})$.
The mixing angles $\theta_{12}$ and $\theta_{23}$ have been precisely measured by solar and atmospheric oscillation experiments. $\theta_{13}$ is also  known from Daya Bay, RENO and T2K experiments. At present, the latest global fit results corresponding to the NH and the IH are given in \cite{GlobalFit20141} (as listed in Tab.\ref{tab.paravalue}).
    \begin{table}[htdp]
    \caption{Neutrino oscillation parameters for NH and IH respectively.}
    \begin{center}
    \begin{tabular}{c|c|c}
    \hline\hline
    parameter & NH & IH
    \\
    \hline
    $\sin^2\theta_{12}/10^{-1}$ & $3.23\pm 0.16$ & $3.23\pm0.16$
    \\
    $\sin^2\theta_{23}/10^{-1}$ & $5.67^{+0.32}_{-1.28}$ & $5.73^{+0.25}_{-0.43}$
    \\
    $\sin^2\theta_{13}/10^{-2}$ & $2.34\pm0.20$ & $2.40\pm0.19$
    \\
    \hline
    $\Delta m_{21}^2[{10^{-5}\text{eV}^2}]$ & $ 7.6^{+0.19}_{-0.18}$ & $ 7.6^{+0.19}_{-0.18}$
    \\
    $|\Delta m_{31}^2|[{10^{-3}\text{eV}^2}]$ & $ 2.48^{+0.05}_{-0.07}$ & $ 2.38^{+0.05}_{-0.06}$
    \\
    \hline\hline
    \end{tabular}
    \end{center}
    \label{tab.paravalue}
    \end{table}%

However, the leptonic CP violation parameter is still unknown. With the increasingly precise measurement capabilities of upcoming neutrino experiments, the effect of the leptonic CP phase must be considered.
When the matter effect must be considered for long-distance detection experiments in which the neutrinos pass through the Earth, the oscillation probability becomes
    \begin{eqnarray}
        &&P_{\nu_\mu\rightarrow\nu_e}\simeq\sin^2\theta_{23}\sin^22\theta_{13}\frac{\sin^2[(1-A)\Delta]}{(1-A)^2}
        \nonumber\\
        &&~~~~+\alpha\sin2\theta_{13}\sin2\theta_{12}\sin2\theta_{23}\cos\theta_{13}\frac{\sin[(1-A)\Delta]\sin [A\Delta]}{A(1-A)}\cos(\delta_{CP}+\Delta)
        \nonumber\\
        &&~~~~+\alpha^2\cos^2\theta_{23}\sin^22\theta_{12}\frac{\sin^2[A\Delta]}{A^2}
    \label{eq.MattEff}
    \end{eqnarray}
with $A=2E V/\Delta m^2_{31}$, {$\alpha=\Delta m_{21}^2/\Delta m_{31}^2$} and $\Delta=\Delta m^2_{31}L/(4E)$ \cite{MatterEff,WPneutrino}. Here, $E$ and $L$ respect the neutrino beam energy and the distance between the source and the detector, respectively.

With the aid of the error transfer formula, the standard deviation of $\Delta\delta_{CP}$ can be calculated using Eq.(\ref{eq.MattEff}), such that it depends on the errors of all phenomenological parameters, including the beam energy $E$ and the distance $L$, as follows:
\begin{eqnarray}
    \left[\delta(\delta_{CP})\right]^2&=&F_1\left[\delta(\sin^2\theta_{12})\right]^2
        +F_2\left[\delta(\sin^2\theta_{13})\right]^2
        +F_3\left[\delta(\sin^2\theta_{23})\right]^2
        \nonumber\\
        &&+F_4\left[\delta(\Delta m_{31}^2)\right]^2
        +F_5\left[\delta(\Delta m_{21}^2)\right]^2
        +F_6\left[\delta E\right]^2
\end{eqnarray}
with the abbreviated function $F_i=F_i(\theta,\Delta m^2, E, L,\delta_{CP})$.
The errors on the probability $P_{\nu_\mu\rightarrow\nu_e}$ and the distance $L$ are treaded ideally.

\begin{figure}[htbp]
\begin{center}
  \includegraphics[scale=0.6]{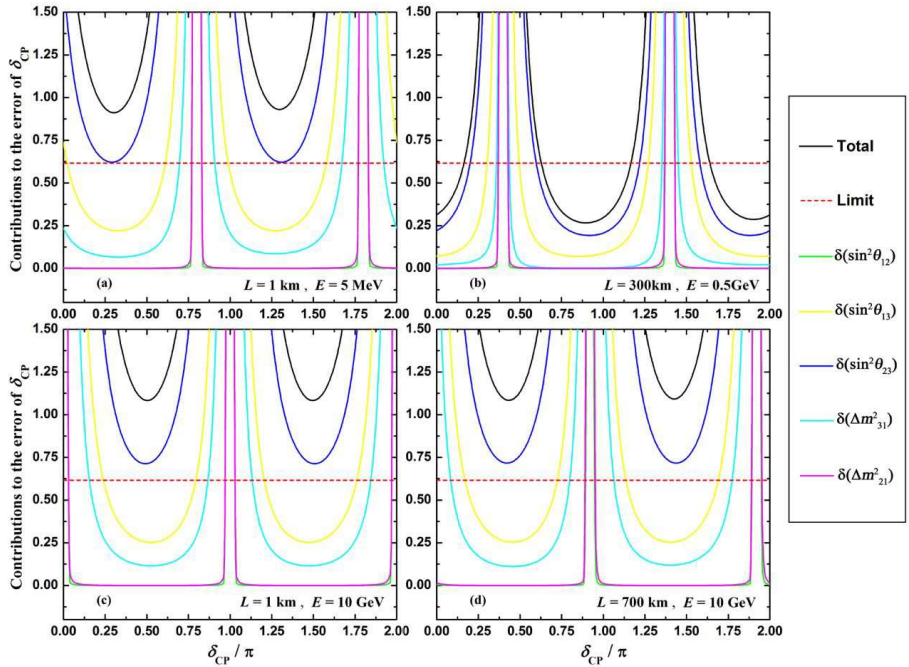}
\caption{Contributions to the error of leptonic CP phase from phenomenological parameter measurements and detector performances at fixed beam energy and detecting distance (a) $L=1$km and $E=5$MeV; (b)$L=300$km and $E=0.5$GeV; (c) $L=1$km and $E=10$GeV; (d)$L=700$km and $E=10$GeV. Red dotted line labels statistical limit $\Delta\delta_{CP}=\pi/4$.}
\label{fig.DeltaCP}
\end{center}
\end{figure}

In Fig. \ref{fig.DeltaCP}, the contributions to $\Delta\delta_{CP}$ for various experimental scenarios are analyzed on a $2\pi$ cycle of $\delta_{CP}$ for select detector parameters, $E$ and $L$. For convenience, the energy resolution is set to zero.

\begin{figure}[htbp]
\begin{center}
  \includegraphics[scale=0.4]{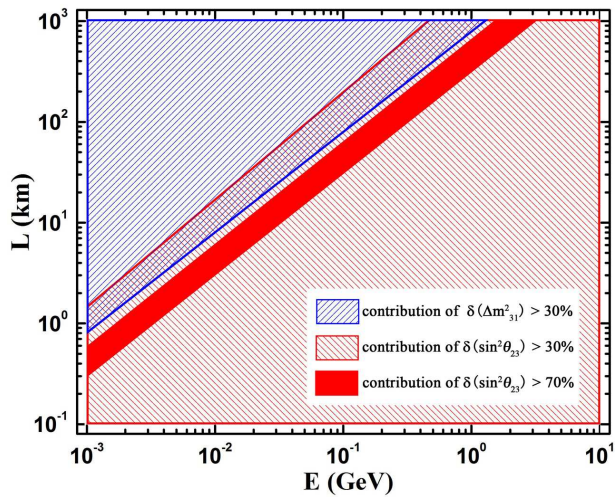}
\caption{Dominant contributions to $\Delta\delta_{CP}$ from two errors of $\Delta m_{31}^2$ and $\Delta\sin^2\theta_{23}$.}
\label{fig.dominant}
\end{center}
\end{figure}

The deviation of the CP phase directly affects the means of its measured values in terms of its statistical distribution. To obtain a separable $\delta_{CP}$ on a $2\pi$ cycle, the maximum of the standard deviation must be limited such that  $\Delta\delta_{CP}=\pi/4$, which corresponds to $95\%$ C.L. ($2\sigma$) on a half of cycle. In Fig. \ref{fig.DeltaCP}, a total deviation of the CP phase of below $\pi/4$ means that $\delta_{CP}$ can be distinguished when it is  locateed in the corresponding interval.  Otherwise, it will be  statistically uncertain.
Fig. \ref{fig.DeltaCP}  exhibits two notable characteristics. The first  is the existence of singular points.
There are always two singular points where $\Delta\delta_{CP}$ tends toward infinity.
They originate from the vanishing sine function in the denominator
    \begin{eqnarray}
    \sin(\delta_{CP}+\Delta)=0
    \label{eq.singularpoint}
    \end{eqnarray}
that occurs when differentiating the second term of Eq.(\ref{eq.MattEff})
{$$\delta[\cos(\delta_{CP}+\Delta)]=\sin(\delta_{CP}+\Delta)\delta(\delta_{CP}).$$}
Obviously, the distance between these two singular points is $\pi$.
The second notable characteristics of the figure is the variability in which parameter's error dominates the uncertainty, which differ for different detection parameters.
The regions of the $E$-$L$ plane corresponding to parameter error contributions of more than 30\% (and 70\%) are depicted in Fig. \ref{fig.dominant}.

\begin{figure}[htbp]
\begin{center}
  \includegraphics[scale=0.4]{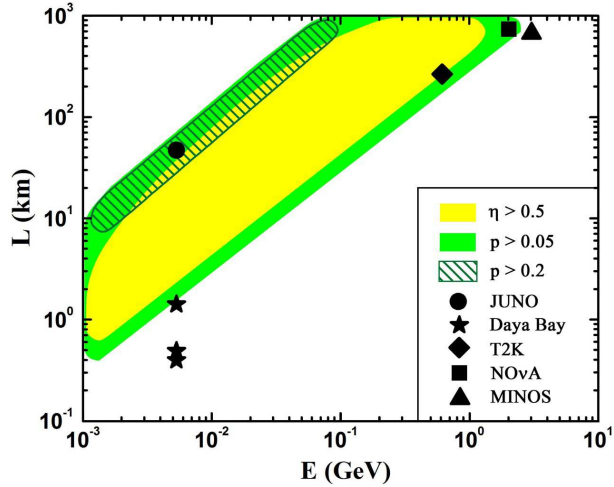}
\caption{Two competitive requirements to $E$ and $L$: significant probability vs. suppressed error of $\delta_{CP}$. }
\label{fig.diffreq}
\end{center}
\end{figure}

From the above results, it is apparent that there are two competing requirements on the beam energy $E$ and distance $L$ for suppressing the error on the leptonic CP violation and increasing the transformation probability. Fig. \ref{fig.diffreq} illustrates the corresponding different regions in the plane of $E$ and $L$. It is  possible to satisfy both requirements by choosing $E$ and $L$ from an area of  intersection.

\begin{figure}[htbp]
\begin{center}
  \includegraphics[scale=0.4]{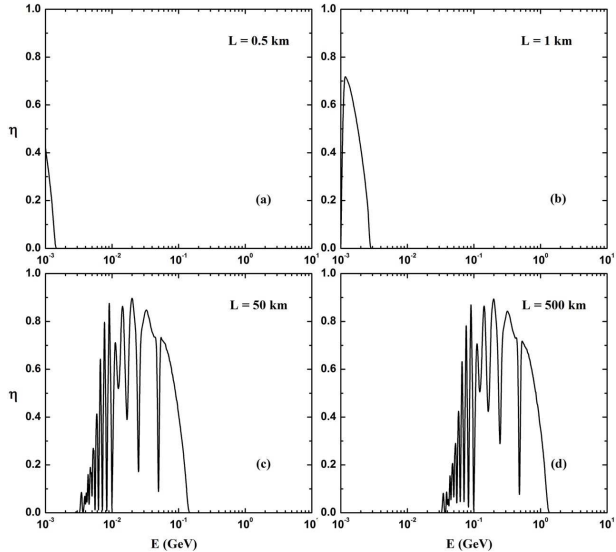}
\caption{Efficiency factor vs. beam energy at fixed distance $L$ (1) $L=0.5km$; (2) $L=1km$; (3) $L=50km$; (4) $L=500km$.}
\label{fig.eta}
\end{center}
\end{figure}

A matter of concern is how to describe the discrimination powers of a detector with respect to an arbitrary leptonic CP violation on a $2\pi$ cycle.
One can introduce an efficiency factor $\eta$ to characterise the effect:
\begin{eqnarray*}
    \eta(E,L)=\frac{\text{the total of intervals below } \Delta\delta_{CP}=\pi/4 \text{on a circle}}{2\pi}.
\end{eqnarray*}
A larger value of $\eta$ indicates a larger range in which $\delta_{CP}$ can be determined. A vanishing $\eta$ indicates that the CP violating phase is indistinguishable CP violation on a $2\pi$ cycle.
When $\eta$ tends toward $1$,  the CP violatingphase can be statistically determined regardless of its measured value. Because of the singular points, it is not possible to achieve an efficiency factor $\eta$ that is equal to 1, even for a detector with perfect energy resolution $\Delta E=0$ that is located at the optimum distance from the source.
Fig. \ref{fig.eta} shows that the efficiency factor can be determined based on the beam energy $E$ and the distance $L$ of a detector. In the low energy region, the efficiency factor $\eta$ appears to oscillate because of the  bottom-lifting effect when $E$ and $L$ depart from their optimum values. As the flat bottom of the curve approaches the  constraint line $\Delta\delta_{CP}=\pi/4$, the interval below the constraint line will vary rapidly.

\begin{figure}[htbp]
\begin{center}
  \includegraphics[scale=0.25]{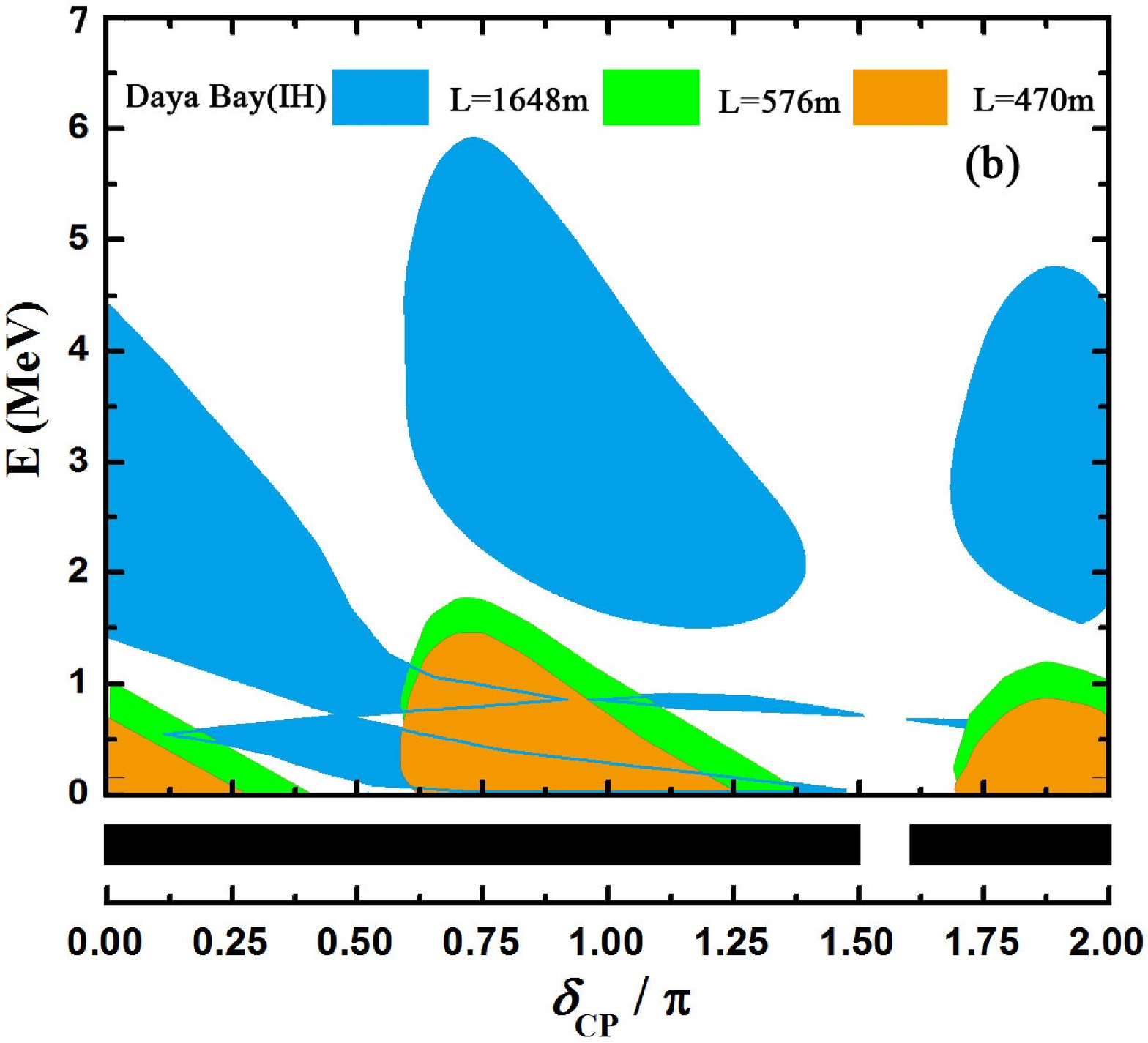}
  \includegraphics[scale=0.25]{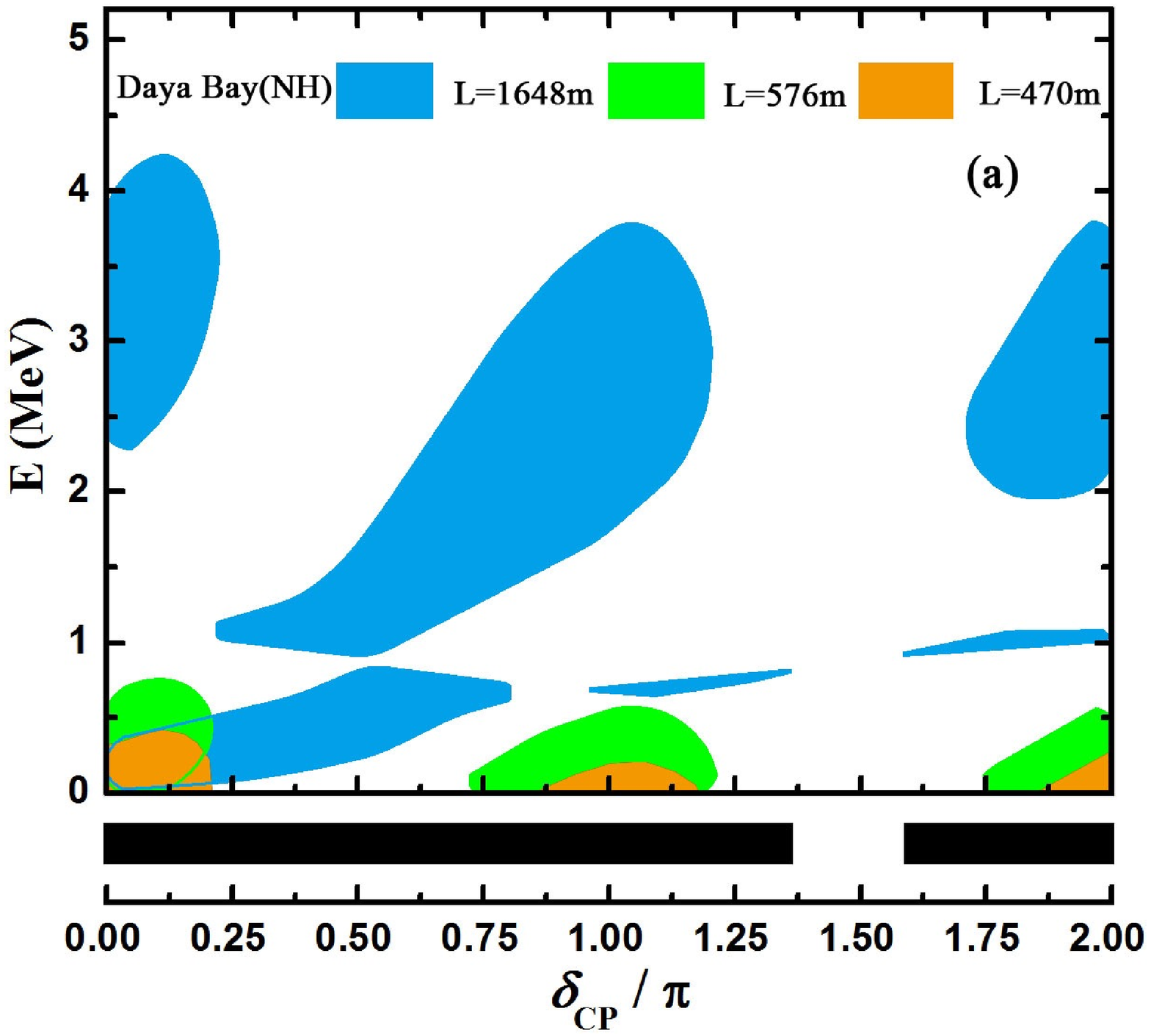}
  \\
  \includegraphics[scale=0.25]{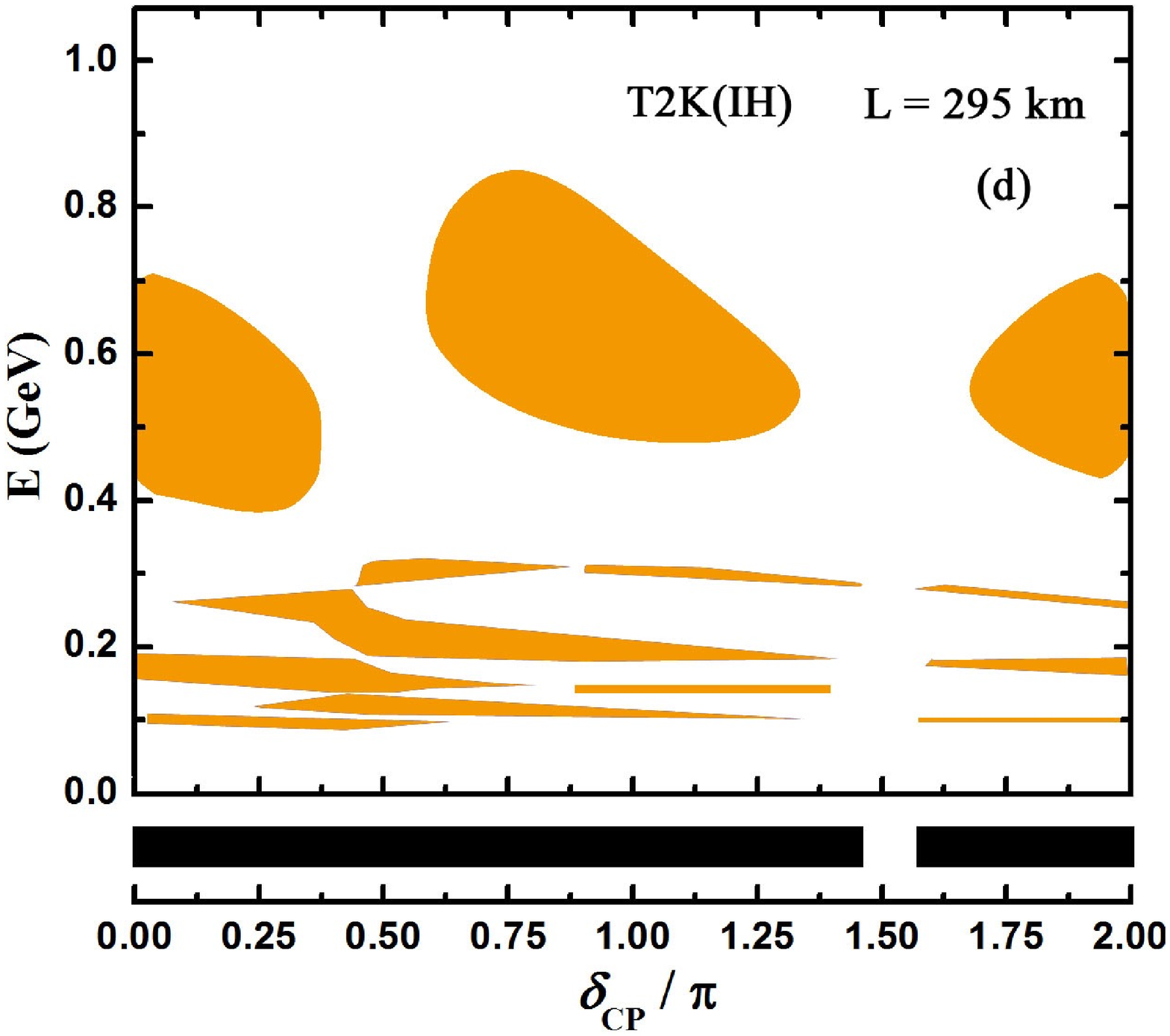}
  \includegraphics[scale=0.25]{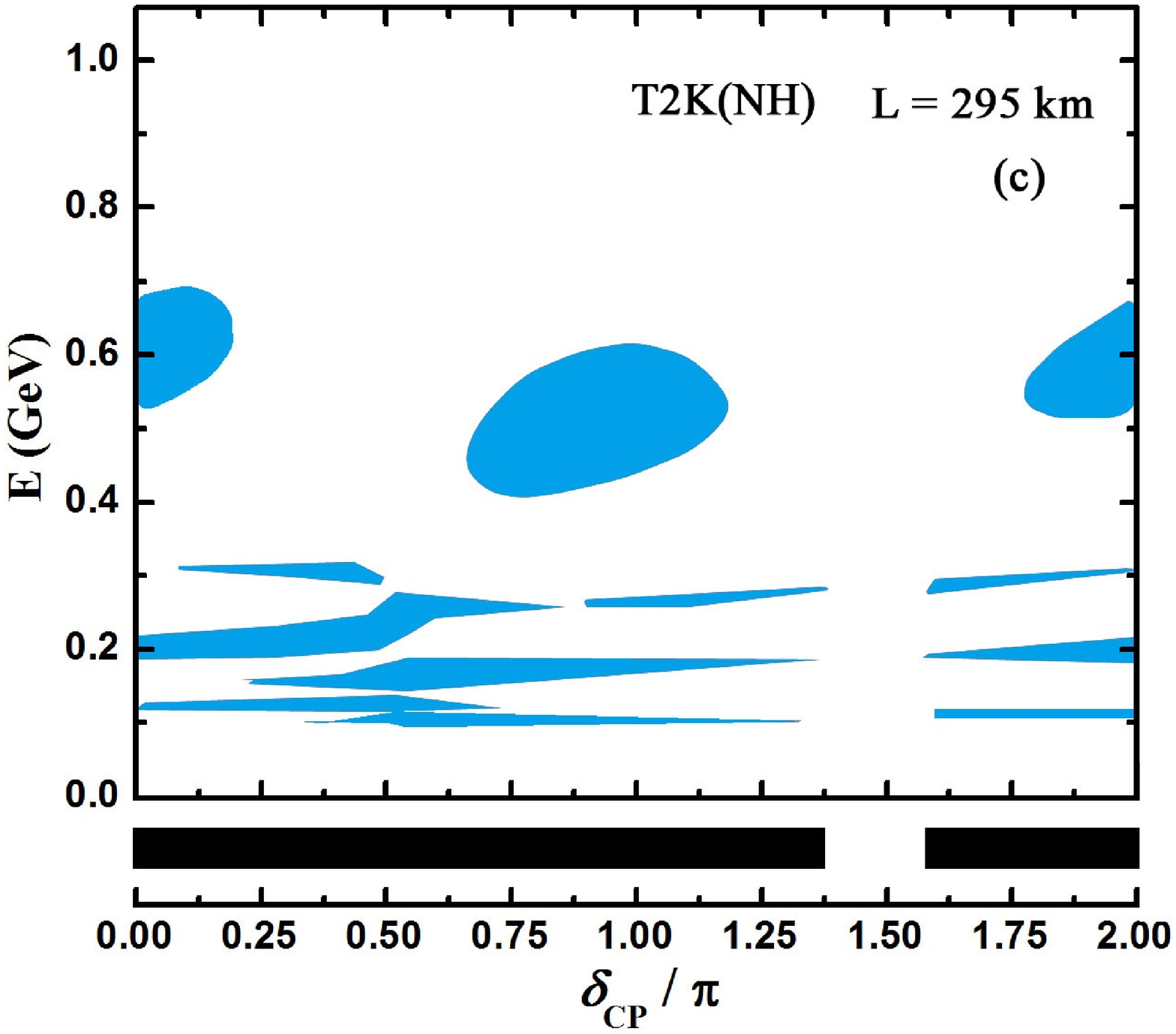}
  \\
  \includegraphics[scale=0.25]{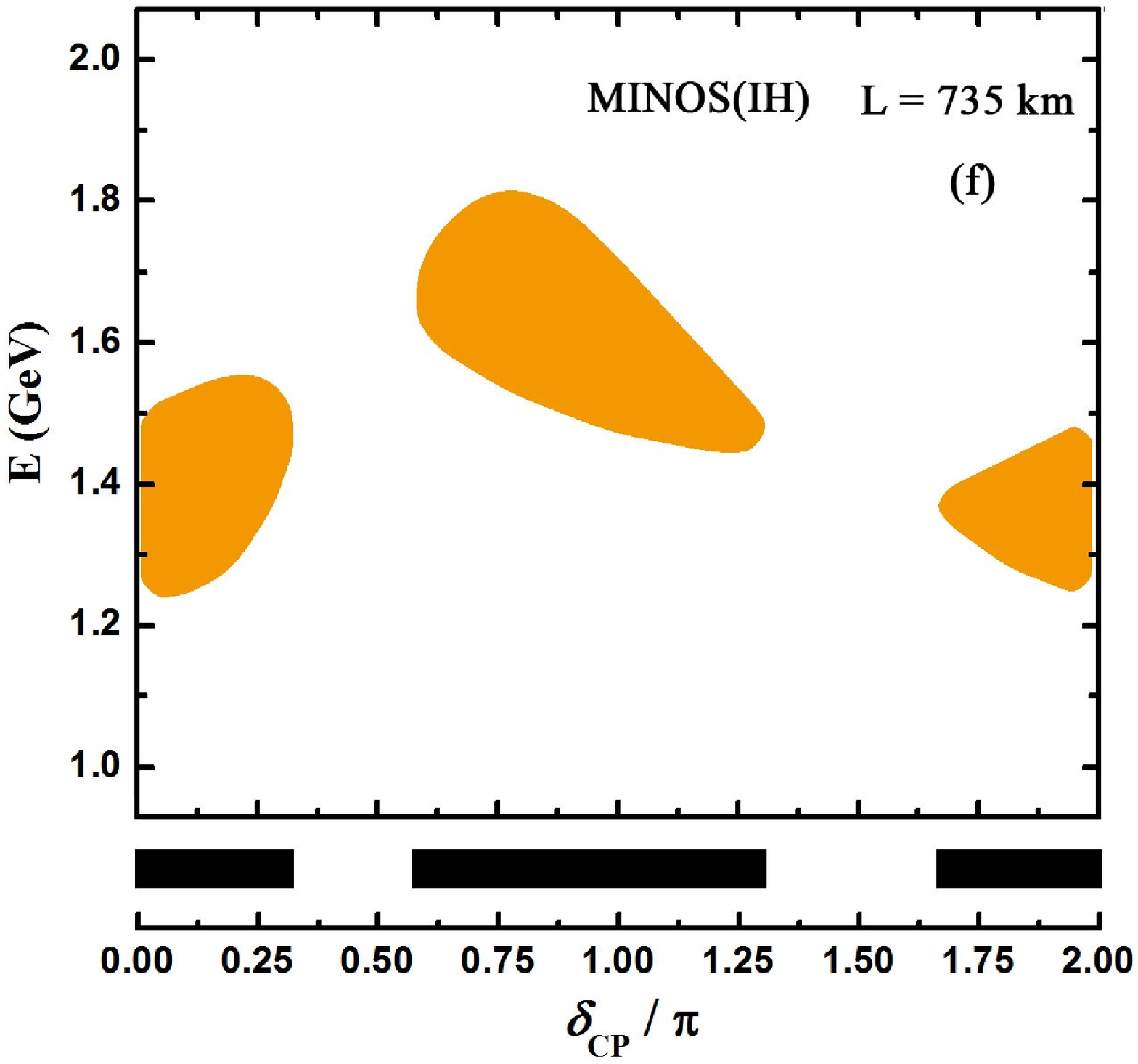}
  \includegraphics[scale=0.25]{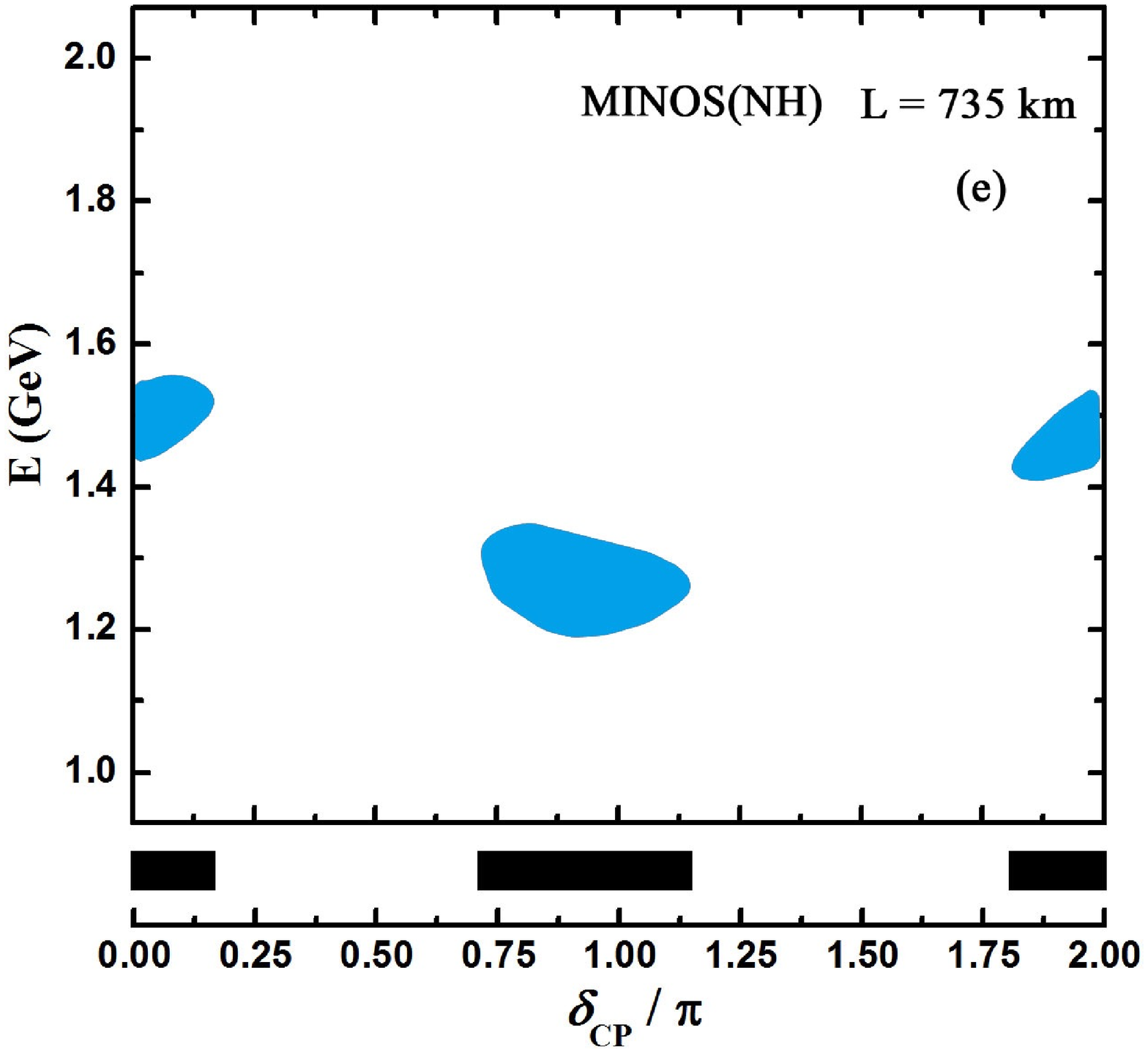}
  \\
  \includegraphics[scale=0.25]{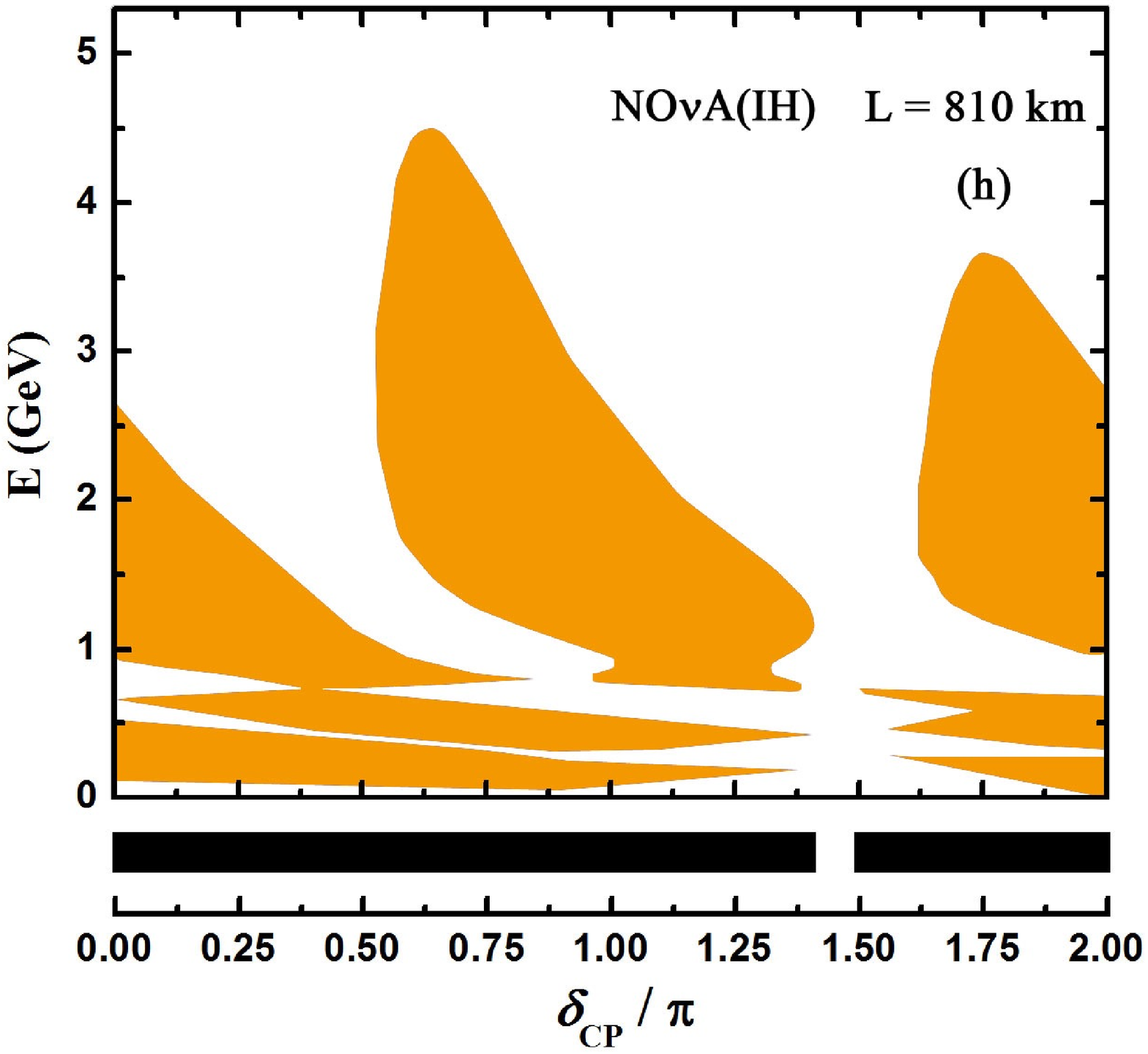}
  \includegraphics[scale=0.25]{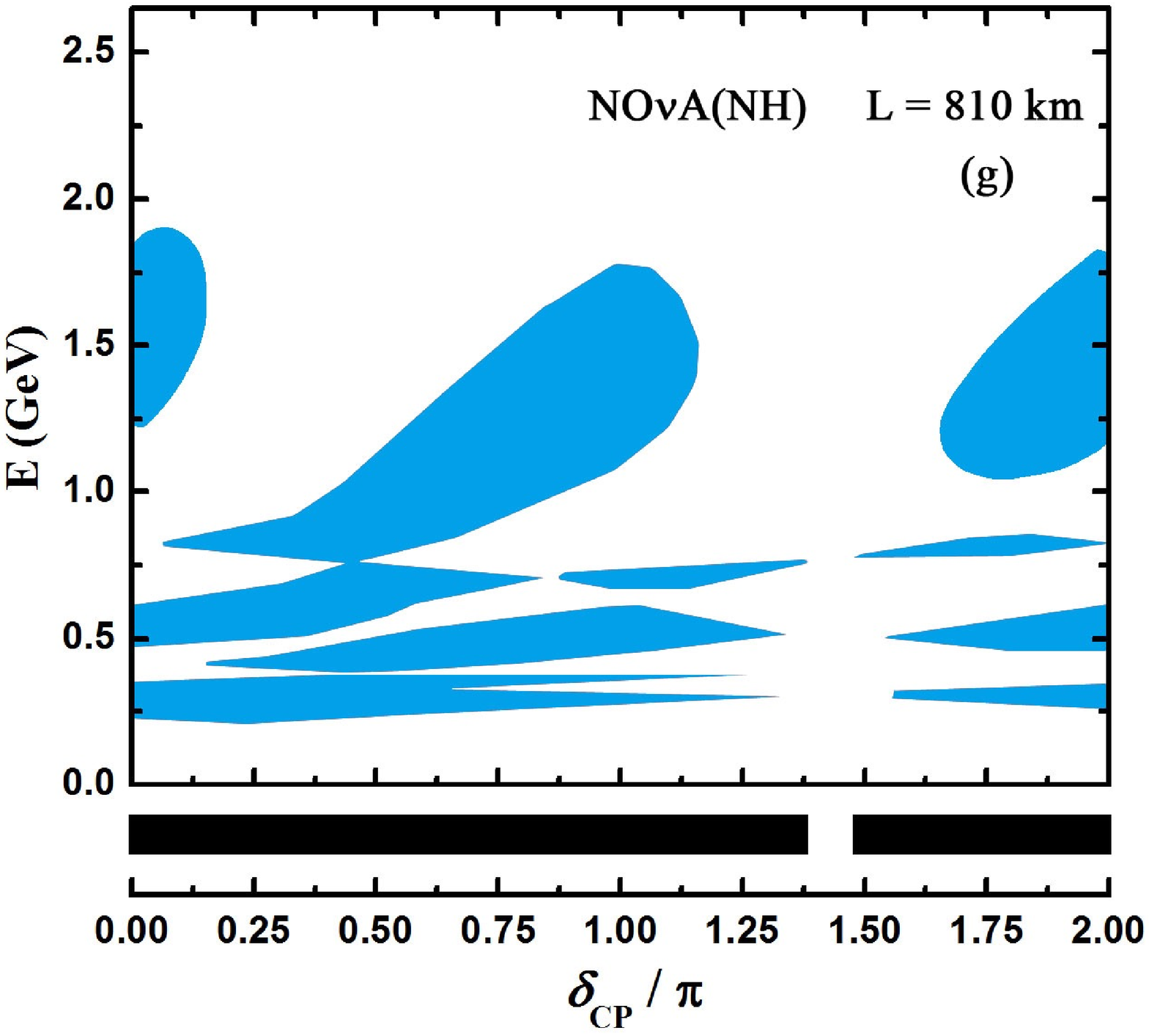}
\caption{Complexed efficiency factors of neutrino experiments for NH and IH. Black segments at the bottom of each subfigures show all possible distinguishable region of $\delta_{CP}$.
Energy resolutions of detectors has been considered actually.}
\label{fig.cometa}
\end{center}
\end{figure}

In practical terms, neutrino detection experiments always span a finite  energy region, and the effect of this range of investigated energies should be expressed in terms of a combined efficiency factor $\eta^*$  that represents the union of all distinguishable intervals for energies $E_i$ running over all possible detection regions:
\begin{eqnarray}
\eta^*=\bigcup_{E_i}\left(\text{intervals below }\Delta_{CP}=\pi/4\text{ at fixed }E_i\right).
\label{eq.ComplexEta}
\end{eqnarray}
In Fig. \ref{fig.cometa}, the combined efficiency factors of several major neutrino experiments are plotted. The efficiency factor at a given  energy $E_i$ can be read out from the sum of the horizontal segments. The projection onto the $\delta_{CP}$ axis of all areas for a given detector represents its maximum distinguishable range.
We find that, at present, there is a blind zone around $\delta_{CP}=3\pi/3$. If the leptonic CP phase is located in this blind zone, then these neutrino experiments will not be able to determine it statistically. To fill in this zone, a new detector must be considered.

In summary, we has investigated the properties of the error on the leptonic CP violation under a proposed statistical constraint that requires $\delta_{CP}$ to be less than $\pi/4$ for its value to be distinguishale on a cycle. In particular, the energy and distance for the neutrino detector were  considered. An efficiency factor $\eta$, was defined to describe the discrimination power of a detector for CP violation. The combined $\eta^*$ values for major neutrino experiments were calculated, and these values provide a clear picture of the distinguishable range of $\delta_{CP}$. In the future, the $\eta^*$ value will be studied further when the relevant phenomenological parameters can be measured more exactly.

\section*{acknowledgement}
We thank our  colleague, Dr. Rong Li for helpful discussions.


\end{document}